\documentstyle[11pt,newpasp,twoside,epsf]{article}
\def\edcomment#1{\iffalse\marginpar{\raggedright\sl#1\/}\else\relax\fi}
\reversemarginpar

\begin{document}
\title{The Extended Tidal Tails of Palomar 5: \\ \smallskip
Tracers of the Galactic Potential}
 \author{Michael Odenkirchen, Eva K.\ Grebel, Hans-Walter Rix}
\affil{MPI for Astronomy, K\"onigstuhl 17, D-69117 Heidelberg, Germany}
\author{Walter Dehnen}
\affil{Ast.\ Inst.\ Potsdam, An der Sternwarte 16,
D-14482 Potsdam, Germany}
\author{Heidi Jo Newberg}
\affil{Rensselaer Polytechnic Institute, 110 Eight St., Troy, NY 12180, USA}
\author{Constance M.\ Rockosi}
\affil{University of Washington, Box 951580, Seattle, WA 98195, USA}
\author{Brian Yanny}
\affil{Fermilab, P.O.\ Box 500, Batavia, IL 60510, USA}

\begin{abstract}
We detected extended, curved stellar tidal tails
emanating from the sparse, disrupting halo
globular cluster Pal 5, which cover $10^{\circ}$
on the sky.  These streams allow us to infer the orbit of Pal 5
and to ultimately constrain the Galactic potential at its location.
\end{abstract}

\section{Palomar 5: A Globular Cluster Torn Apart by the Milky Way}

Palomar~5 is an extraordinarily sparse globular cluster 
in the outer halo of the Milky Way, at a distance of 23~kpc from the Sun 
(Fig.1). 
Its peculiar properties (e.g., very low mass, large core, relatively flat 
luminosity function) fostered the idea that this cluster might be a likely 
victim of disruptive Galactic tides.

Using deep multi-color photometry from the Sloan Digital Sky Survey (SDSS; 
York et al.\ 2000, Gunn et al.\ 1998) 
we found unambiguous, direct evidence for the suspected 
tidal disruption of Pal\,5 (Oden\-kir\-chen et al.\ 2001; 
Rockosi et al.\ 2002):  For the first time, two massive tails
of stellar debris with well-defined S-shape geometry were 
detected, emanating in opposite directions from the cluster. 

As the SDSS is scanning more and more of the sky we have now extended our
search over an area of $\sim$87 deg$^2$. 
Contaminating objects were removed by eliminating extended sources and by 
applying an optimized smooth color-magnitude-dependent weighting function. 
This optimized weighting enhances the density contrast between cluster 
and field stars by almost a factor of 20 and provides a least-squares 
solution for the spatial distribution of the cluster population. 
The resulting surface density map of Pal\,5 stars is shown in Fig.~2. 

\begin{figure}[ht]
\plotone{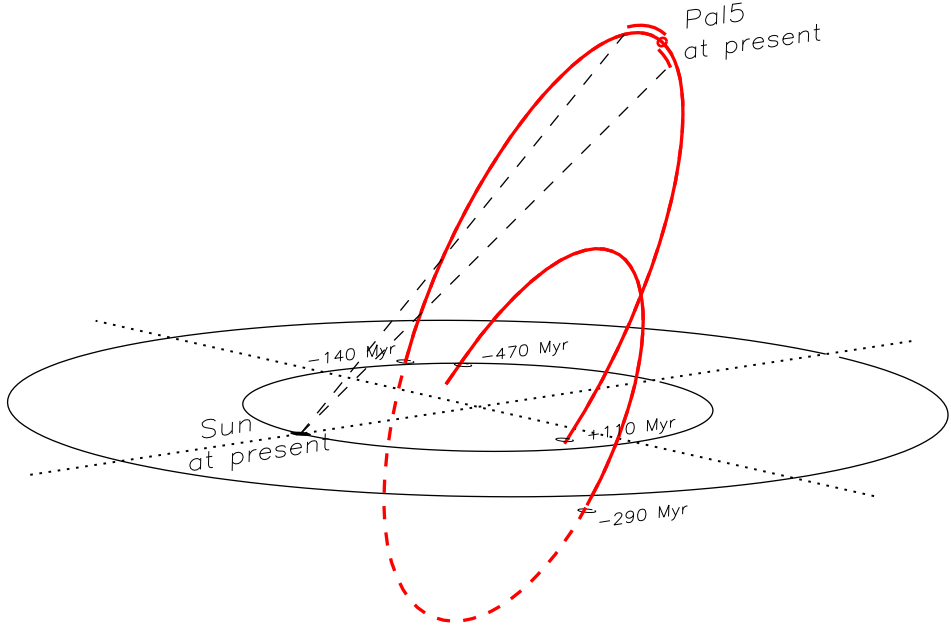}
\caption{
The present location of Palomar 5 
in the Milky Way and its inferred plunging 
Galactic orbit (fat solid and dashed line).
The approximate times of past and future disk passages 
are indicated. 
}
\end{figure}

\section{A Narrow, Curved, 10 Degree Stream of Debris}

We find that the tidal tails extend over an arc of at least 
10$^\circ$ on the sky and form a narrow stream with a FWHM of only $18'$. 
This corresponds to a projected length of $\simeq$~4~kpc in space, and a 
projected FWHM of 120 pc. 
The northern tail is visible out to 6.5$^\circ$ from the cluster.
The southern tail is traced over 3.5$^\circ$ but probably continues 
beyond the border of the currently available field (Fig.\ 2). 
The stellar mass in the tails adds up to 1.2 times the mass of stars in 
the cluster, i.e., the tails contain more mass than what is left in the 
cluster.
Pal\,5 thus presents a text-book example of a tidally 
disrupting globular cluster. It is so far the only known stellar system 
besides the Sagittarius dwarf galaxy that demonstrates 
the formation of a halo stream within the Milky Way. 

The tails have a clumpy structure (Fig.\ 2). This implies that 
the mass loss has been episodic, and suggests that it was triggered by 
disk and/or bulge shocks.  Indeed Pal\,5 passes through the Galactic 
disks at intervals of a few 100 Myr (Fig.\ 1). 
In Fig.~3 we present the radial profile of the stellar surface density 
(i.e., the azimuthally averaged surface density as a function of distance 
from the cluster center) from the core of the cluster out to the current 
end points of the tails. 
The profile shows a characteristic break near the cluster's 
tidal radius at about $16'$. 
Inside this radius, the profile decreases approximately
like $r^{-3}$. Beyond this limit the profile is flatter and  
approximately follows an $r^{-1.5}$ power law. 
The overall decrease thus differs from 
a simple $r^{-1}$ power law that would result from a constant linear 
density along the tails.  

\begin{figure}[ht]
\plotone{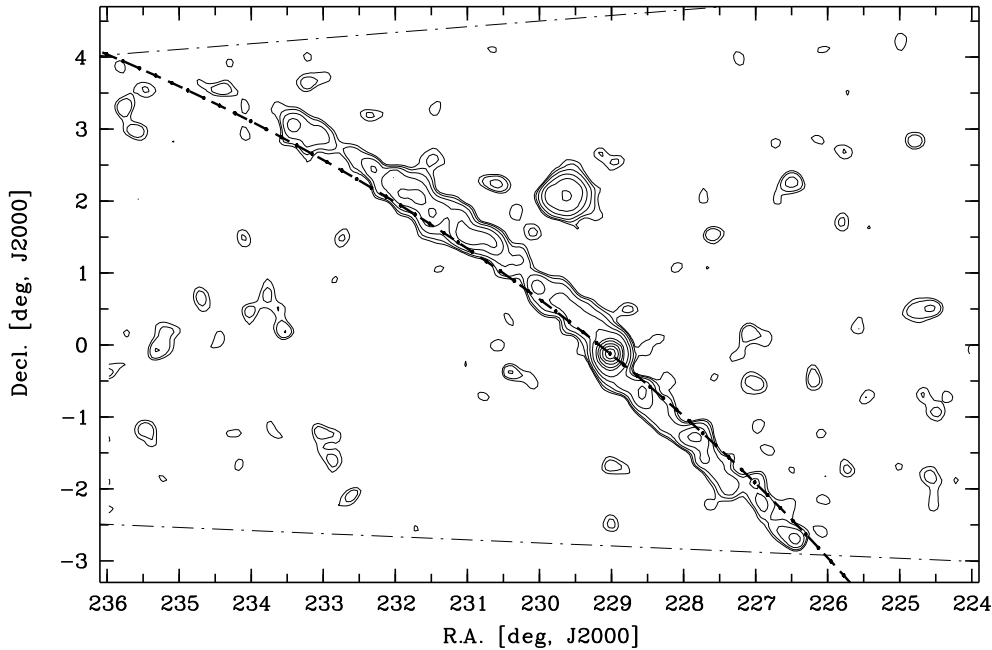}
\caption{
Contour map of the surface density of Pal\,5 stars 
derived by weighted star counts of SDSS point sources. 
The fat dashed line shows the best-fit local orbit of the cluster.
The curved dash-dotted lines indicate the border of the current
SDSS scan region.
}
\end{figure}

\section{Clues on the Cluster's Orbit and Mass Loss Rate}

Location and curvature of
the tails are direct tracers of the cluster's Galactic motion  
and hence provide unique information about the orbit of Pal\,5. 
The best-fit local orbit is shown as dashed line in Fig.~2.
The direction of the cluster's motion is determined by its orientation
with respect to the Galactic center:  The southern, leading tail
and the northern, trailing tail indicate that Pal\,5 is on a prograde
orbit.
Using a standard three-component model for the Galactic potential
we infer that Pal\,5 is observed close to the tails' maximum distance from 
the Galactic disk and has recently passed through apogalacticon 
(implying that the tidal stream is currently relatively dense). 
In about 100 Myr the cluster will cross the disk at a distance of 
only 7~kpc from the Galactic center (see Fig.~1). 
This will produce a strong tidal shock that might lead to complete 
disruption.

The amount by which the tails are offset from the orbit of the 
cluster is directly related to the velocity at which the tidal 
debris drifts away from the cluster. The observed mean offset 
(about 75 pc in projection), the 
parameters of our model orbit, and the total amount of  
stellar mass seen in the tails lead to an estimate of the
mean mass loss rate of about $5\,M_\odot$ Myr$^{-1}$. 
Assuming this rate to be more or less constant (as suggested by  
numerical simulations) we conclude that 10 Gyr ago Pal\,5 may have 
had a mass of about $5 \cdot 10^4 M_\odot$. This is about ten 
times as much as it has today, but still considerably less than 
the mass of an average present-day Galactic globular cluster.

\begin{figure}[ht]
\plotone{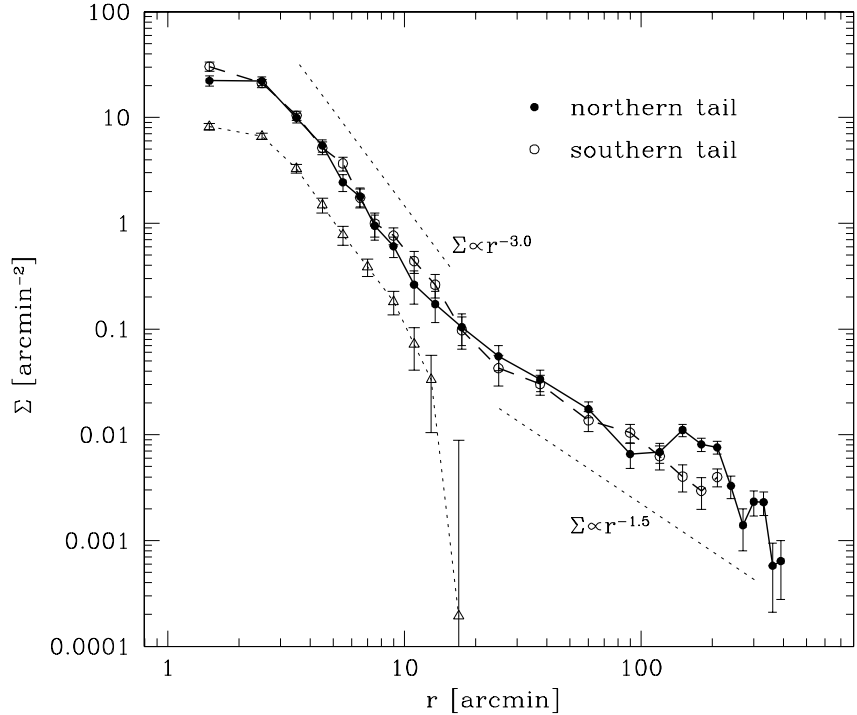}
\caption{
Radial profile of the stellar surface density in the 
cluster and tails from number counts in sectors of concentric 
rings. For comparison the leftmost data points show the cluster profile 
from counts in sectors perpendicular to the tails 
(arbitrarily shifted by $-1$ in log $\Sigma$).
}
\end{figure}

The current data allow us to predict the tangential
velocity of Pal\,5 as a function of the parameters of the Galactic
potential.  GAIA and SIM will allow us to accurately measure the
velocities and proper motions of stars in the cluster and in the 
extended tails, fully characterizing the kinematics of the
stream independent of any galactic model.  In return, these kinematics
impose strong constraints on the Galactic potential at the location of 
the cluster.

\acknowledgments

Funding for the creation and distribution of the SDSS Archive has been 
provided by the Alfred P. Sloan Foundation, the Participating Institutions, 
the National Aeronautics and Space Administration, the National Science 
Foundation, the U.S. Department of Energy, the Japanese Monbukagakusho, 
and the Max Planck Society. The SDSS website is http://www.sdss.org/.

\end{document}